\documentclass[a4paper,11pt]{article}

\usepackage{fullpage}

\usepackage{algorithmic}
\usepackage{algorithm}
\usepackage{url}
\usepackage{graphicx}
\usepackage{color}
\usepackage{listings}
\usepackage{subfigure}
\usepackage{authblk}

\usepackage{tikz}
\usetikzlibrary{shapes}

\title{Configurable Strategies for Work-stealing\thanks{The
research leading to these results was partially funded by the
European Union Seventh Framework Programme (FP7/2007-2013)
under grant agreement no.\ 248481 (PEPPHER Project, \protect\url{www.peppher.eu}). A short poster summary of this paper was presented at the 18th ACM PPoPP 2013 conference~\cite{ppopp13}.}
}

\author[1]{Martin Wimmer}
\author[2]{Daniel Cederman}
\author[1]{Jesper Larsson Tr{\"a}ff}
\author[2]{Philippas Tsigas}
\affil[1]{Faculty of Informatics\\Vienna University of Technology\\
Favoritenstrasse 16, 1040 Vienna, Austria\\ \texttt{\{wimmer,traff\}@par.tuwien.ac.at}}
\affil[2]{Computer Science and Engineering\\Chalmers University of Technology\\
412 96 G\"oteborg, Sweden\\ \texttt{\{cederman,tsigas\}@chalmers.se}}

\begin{document}

\maketitle

\begin{abstract}
Work-stealing systems are typically oblivious to the nature of the
tasks they are scheduling. For instance, they do not know or take into
account how long a task will take to execute or how many subtasks it
will spawn.  Moreover, the actual task execution order is typically
determined by the underlying task storage data structure, and cannot
be changed.  There are thus possibilities for optimizing task parallel
executions by providing information on specific tasks and their
preferred execution order to the scheduling system.

We introduce \emph{scheduling strategies} to enable applications to
dynamically provide hints to the task-scheduling system on the nature
of specific tasks.  Scheduling strategies can be used to independently
control both local task execution order as well as steal order. In
contrast to conventional scheduling \emph{policies} that are normally
global in scope, strategies allow the scheduler to apply optimizations
on individual tasks.  This flexibility greatly improves composability
as it allows the scheduler to apply different, specific scheduling
choices for different parts of applications simultaneously.  We
present a number of benchmarks that highlight diverse, beneficial
effects that can be achieved with scheduling strategies. Some
benchmarks (branch-and-bound, single-source shortest path) show that
prioritization of tasks can reduce the total amount of work required
compared to standard work-stealing execution order. For other
benchmarks (triangle strip generation) qualitatively better results
can be achieved in shorter time. We also demonstrate that strategies
are composable.  Other optimizations, such as dynamic merging of tasks
or stealing of half the work, instead of half the tasks, are also
shown to improve performance.  Compositionality of strategies is
demonstrated by examples that combine different strategies, both
within the same kernel (prefix sum) as well as when scheduling
multiple kernels (prefix sum and unbalanced tree search).
\end{abstract}

\section{Introduction}

Work-stealing is a popular way to schedule parallel work-loads of
independent tasks~\cite{BlumofeLeiserson99} and is used by well-known
frameworks such as
Cilk~\cite{BlumofeJoergKuszmaulLeisersonRandallZhou96},
Cilk\verb!++!~\cite{Leiserson10}, Intel Threading Building
Blocks~\cite{KukanovVoss07} and X10~\cite{Charles05}.  Whenever new
tasks are created within an application, they are stored in a local
queue owned by a specific thread. When a thread runs out of work in
its own queue, it tries to steal work out of the queues of other
threads.

Standard work-stealing schedulers are oblivious to most properties of
individual tasks and treat tasks equally.  The execution order of
tasks owned by a thread is instead indirectly determined by the
data structure used for storing the tasks.  A commonly used
data structure for work-stealing systems is the lock-free
work-stealing deque by Arora et al.~\cite{AroraBlumofePlaxton01} that
is optimized for local accesses.  The owning thread accesses the queue
from one end, whereas stealing threads access it from the other.  The
order of tasks in the queue is often a good heuristic for an efficient
execution order of tasks. First, the owning thread accesses the queue
like a stack, which leads to a depth-first execution of tasks and
often results in good performance because of good memory use and
locality~\cite{AcarBlellochBlumofe02}. Second, stealing
threads access the queue from the other end, resulting in a
\emph{first-in-first-out} access pattern. This leads the stealer to
acquire a new task near the root of the task graph, thereby generating
more local work and reducing the number of further steals.

While the
execution order effected by work-stealing deques
is good for some applications, other execution orders are better for
other applications. Search-based algorithms can profit from
prioritization to explore the most promising branches early. Other
algorithms, like prefix-sums (as presented in
Section~\ref{sec:prefix}), can combine two passes on data into one if
the tasks are executed in the right order. Other applications benefit
from a regime which gives priority to tasks that access data already
in the cache~\cite{Weissman98}, and such locality-aware scheduling
regimes are often used~\cite{GuoZhaoCaveSarkar10}. Another type of
regime would prioritize a task depending on how recently it was
previously scheduled~\cite{Squillante93}.  Another common heuristic is
to prioritize tasks on the critical path~\cite{Song09}. Resource
obliviousness has been achieved with a special priority scheduling
scheme~\cite{ColeRamachandran10}.  A variety of task-parallel
application kernels that profit from prioritization is presented by
Lenharth et al.~\cite{pingali2011}. The authors postulate that a
global priority ordering for tasks is often not beneficial for
performance, and that different priority orderings within the same
application/system are required.

In this paper we present \emph{scheduling strategies} as a way of
informing the scheduling system about properties of tasks and as a way
to prioritize tasks. This makes it possible for a work-stealing scheduler
to improve the execution without losing any generality of the
scheduler. In contrast to \emph{scheduling policies}, which are global
in nature, strategies allow specification of those properties at the
level of individual tasks. Strategies in our system are
\emph{composable}. Regardless of which strategies/types of strategies
are combined the behavior of the scheduler is always
well-defined. Different kernels with different strategies can
therefore be combined in a single parallel execution with the same
scheduler.

We define scheduling strategies in Section~\ref{sec:strategies}, and
explain the kinds of optimizations that we can currently support.
Section~\ref{sec:sched} gives glimpse of our strategy-aware
work-stealing scheduler, without presenting, however, the details
about the data structures required to support strategies; these will
be presented separately.  Example application kernels and the
strategies used to improve them are discussed in
Section~\ref{sec:applications}. Corresponding experimental results in
Section~\ref{sec:results} show that considerable performance
improvements can often be achieved.

\section{Strategies}
\label{sec:strategies}

Scheduling strategies is a mechanism to inform a work-stealing
scheduling system about properties of individual tasks in order to
influence and improve the execution. A strategy can be
associated with a task at spawn time. In contrast to scheduling
policies that are global in nature the scope of a scheduling
strategy is an individual task. This allows to influence the scheduler
behavior for a single task without incurring possibly negative effects
for (all) other tasks. Scheduling strategies are composable, and
different strategies can be used in a single task-parallel
execution because there is a well-defined way in which such strategies
interact.

\paragraph{Spawn to call}

For tasks with small granularity, spawn overhead can significantly
influence the total application execution time. On the other hand, too
coarse grained tasks may lead to too little parallelism or less than
optimal load-balancing.  Spawn overhead can be reduced by converting
task spawns to function calls at run-time. This should preferably be
done dynamically, when the scheduler has a large number of unprocessed
tasks in its queues, thereby trading excess parallelism for a
lower scheduler overhead.  We have noticed that this simple heuristic
can lead to a significant performance improvement for applications
with either small variance in task granularity (algorithms on
same-sized blocks) or decreasing task granularities
(divide-and-conquer algorithms).
For other types of algorithms, the heuristic can be problematic since
it is oblivious to task granularity. In the worst-case, high
granularity tasks would be converted to function calls, and low
granularity tasks put into the task queues.  Strategies avoid such
pathologies by allowing for specifying the task granularity within the
strategy.

Strategies make it possible to control the conversion of task spawns
to synchronous function calls based on properties of the task to be
spawned and the state of the system. A \emph{transitive weight} is
associated with tasks to be used as an estimate of the work that will
be generated by a task and its descendants. Below a certain threshold
which can depend dynamically on, e.g., the number of tasks in the
local task queue, the spawn is converted to a function call.  It is
also possible to disable call conversion and this is done by default
in strategies, unless explicitly enabled.

\paragraph{Number of tasks to steal}

For work-stealing systems it is well known that it is usually better
to steal half the work instead of only a single
task~\cite{Berenbrink01}, the advantage being that work available in
one queue quickly disseminates to the whole system.  In standard
work-stealing systems the amount of work incurred by the tasks is not
known (by the scheduler), and stealing half the work is approximated
by stealing half the tasks. In many cases, this approximation is
highly inaccurate.  For example, in many divide-and-conquer algorithms
the amount of work is halved at each spawn.  To steal half the work in
such algorithms it would be sufficient to steal only the task with the
largest amount of work, instead of half the tasks.

The transitive weight associated with tasks can be used to estimate
the work required by each new task and its descendants. This allows
the stealing procedure to terminate as soon as half the work has been
stolen, irrespectively of the number of tasks in the queues.

\paragraph{Priority}

An application specific execution order of tasks can lead to higher
efficiency (performance, memory usage, quality of the results) than a
fixed execution order like \emph{last-in-first-out}.  Strategies can
be used to suggest an execution order to the scheduling system by
giving the user a means to prioritize tasks of the same type.
Prioritization is implemented by a comparison function that takes two
instances of strategies of the same type and determines which should
be preferred over the other.  Since each instance of a strategy is
associated with a single task, the prioritization of an instance leads
to the prioritization of a task.

The prioritized order of tasks is used locally and when stealing
tasks. Global prioritization is not enforced, since this could
compromise locality and scalability of the work-stealing system; this
is, however, dependent on the data structures used for storing tasks
(and subject to current work not described here). Also note that
conversion of spawns to function calls may violate the
prioritization.

\paragraph{Dead tasks}

For certain applications, like search-based algorithms, newly
calculated results can make tasks obsolete so that they do not need to
be executed any more. Strategies allow the user to expose such
\emph{dead tasks} so that they can be removed early and will not be
stolen by other threads.

\paragraph{Locality}

Together with the notion of \emph{place}, which denotes a single
execution unit in a scheduling system and its supporting
data structures, the task prioritization mechanism provided with
strategies can be used to implement many spatial and temporal locality
optimizations.  Typically the number of places equals the number of
processors in a system.  To facilitate locality optimizations, each
place is bound to a specific processor in the system.

Standard work-stealing systems employ a simple locality optimization,
which is surprisingly good for many applications. Each place has its
own queue of tasks that are processed in \emph{last-in-first-out
order}. This increases temporal locality, since newly created tasks
often work on the same data as their parent tasks. The queues of other
places are not touched as long as the queue of the place is not empty,
which reduces interference. When stealing tasks from another place,
they are stolen in \emph{first-in-first-out} order, which means that
tasks with higher temporal locality are not stolen.
For the remainder of this paper we will call the
standard work-stealing prioritization the \emph{LIFO/FIFO strategy}. This
is the default prioritization for tasks in our system.

Some applications may profit from problem specific locality
optimizations. For algorithms with little temporal locality in the default
strategy a completely different execution order for tasks may be better.

All the described cases can be covered using the prioritization
function, which was explained previously.  Each instance of a strategy is
associated with a place specified by the programmer. By default the
place is set to where the associated task was spawned. The implementer
of the prioritization function can query the associated places as well
as the (memory) distance to the place that requests the
prioritization. Thereby, the programmer can create place-specific
orderings, e.g., by prioritizing tasks with a smaller memory
distance. Since each place may see a different order of tasks, this
must be supported by the task storage data structure.

\paragraph{Composability}

A major design goal of scheduling strategies is composability. It
should be possible for different applications or parts of the same
application that use different strategies to run concurrently within
the same, single scheduler. While this is simple to achieve for
properties that are specific for individual tasks, like the
\emph{transitive weight} of tasks, it is much more difficult for
prioritization.

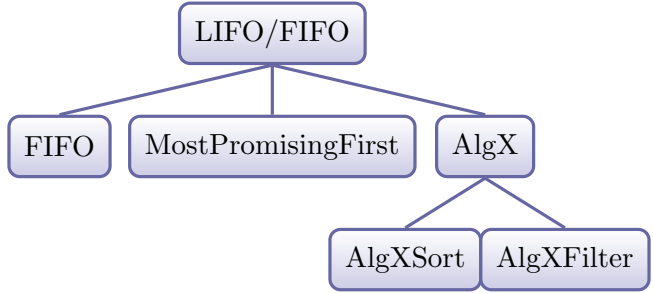
\begin{figure}
\begin{center}
\begin{tikzpicture}[
    grow=down,
    level 1/.style={sibling distance=2.8cm,level distance=1.5cm},
    level 2/.style={sibling distance=2.1cm, level distance=1.5cm},
    edge from parent/.style={very thick,draw=blue!40!black!60},
    edge from parent path={(\tikzparentnode.south) -- (\tikzchildnode.north)},
    every node/.style={text ragged, inner sep=2mm},
    strategy/.style={rectangle, rounded corners, shade, top color=white, minimum height=0.8cm,minimum width=1cm,
    bottom color=blue!50!black!20, draw=blue!40!black!60, very
    thick }
    ]

\node[strategy] {LIFO/FIFO}
    child
    {
        node[strategy] {FIFO}
    }
    child
    {
        node[strategy] {MostPromisingFirst}
    }
    child
    {
        node[strategy] {AlgX}
        child
        {
                    node[strategy] {AlgXSort}
        }
        child
        {
                    node[strategy] {AlgXFilter}
        }
    }
    ;
\end{tikzpicture}
\end{center}
\caption{A hierarchy of scheduling strategies with LIFO/FIFO as the
  base strategy.}
\label{fig:strat_hier}
\end{figure}

We solve the composability problem by imposing a hierarchy of
strategies as shown in Figure~\ref{fig:strat_hier}.  Strategies of
different types are prioritized by the strategy of their lowest common
ancestor. Since the hierarchy has a single root, any two strategies
have a unique common ancestor.  A \emph{LIFO/FIFO strategy}, similar
to the standard work-stealing task order, is the default root
strategy.  The hierarchy allows for different algorithmic kernels in a
single application to use different strategies as indicated in
Figure~\ref{fig:strat_hier}. While some kernels might rely on the base
strategy LIFO/FIFO, another kernel might exhibit better performance
with a FIFO (first-in-first-out) strategy. Search algorithms, on the
other hand are often faster with strategies where the most promising
path is explored first.  More complex, real applications often consist
of different algorithmic kernels.  In Figure~\ref{fig:strat_hier} we
included an algorithm, \emph{AlgX}, which calls both sorting
(AlgXSort) and filtering (AlgXFilter) kernels inside. Both kernels
might require specialized strategies for efficient execution. In
addition, \emph{AlgX} might need to reduce its critical path length by
ordering different calls to sort and filter. This behavior can be
achieved with a common base strategy for both the sorting and the
filtering strategy for \emph{AlgX}.

To impose an absolute ordering on tasks with different type of
strategies, we always let child strategies overrule their
ancestors. For the strategies in Figure~\ref{fig:strat_hier}, this
results in the \emph{FIFO} strategy overruling the \emph{LIFO/FIFO}
strategy. This is done by first grouping all tasks that use the
\emph{FIFO} strategy together and then ordering them in \emph{FIFO}
order. The highest priority task in the group is then compared with
strategies of other types using the \emph{LIFO/FIFO} strategy to
determine the priority of the group as a whole.

\subsection{Implementation of scheduling strategies}

Our implementations are done in an object-oriented, C\verb!++!
framework, called
Pheet~\cite{mtaap13,ppopp13}\footnote{\url{http://www.pheet.org}}. A
scheduling strategy is a class derived from a base strategy class that
implements base functionality required by all strategies, and a
default behavior. Strategies derived from the base strategy can
provide different behavior, for example a different prioritization of
tasks, by overriding the default behavior. The constructor of a
strategy class is allowed to take any kind of parameter the programmer
desires, which allows strategies to act on problem specific
information.  An instance of the strategy class is created and stored
for each spawned task. These objects are then used by the scheduler to
make scheduling decisions for the specific task and to determine the
execution order of the stored tasks.

Algorithm~\ref{lst:strategy} depicts an example implementation of a
strategy that provides depth-first execution for locally spawned
tasks, and a breadth-first execution for tasks created at other
places. It assumes a tree-like algorithm where all tasks in the
subtree will be generated.  The constructor of the strategy stores the
height $h$ of the given task, and sets the \emph{transitive weight} of
the task to $2^h$.  We assume here that the height can never exceed
the number of bits in a long integer, so that no integer overflow will
occur.

To enable conversion of task spawns to function calls, which is
disabled by default, the \texttt{allow\_call\_conversion} method has
to be redefined to return true. More complex strategies may only allow
call conversion for some tasks, by dynamically deciding on the return
value.

The \texttt{prioritize} method determines the execution order of
tasks. It takes a reference to a second strategy object of the same
type as parameter and should return true if the task associated with
the current instance of the strategy should be executed before the
other task, and false otherwise. Algorithm~\ref{lst:strategy}
implements different behaviors depending on whether a task was spawned
in the same place or not. Tasks spawned in the same place are
prioritized for locality reasons and are executed in depth-first
order. All tasks spawned in other places (e.g. stolen tasks in a
work-stealing scheduler) are executed in breadth-first order to
increase the amount of locally spawned work.

\begin{algorithm}
\begin{lstlisting}[mathescape=true,columns=flexible,escapechar=^]
class DepthFirstStrategy : public Environment::BaseStrategy {
public:
	DepthFirstStrategy(int depth, int max_depth)
	:depth(depth)
	{ 	
		// Work is exponential in the height <= max_depth-depth
		set_transitive_weight((long)1 << (max_depth - depth));
	}

	bool allow_call_conversion() const {
		return true;
	}
	
	bool prioritize(DepthFirstStrategy& other) {
		if(this->place == Environment::get_place()) {
			// This task has been spawned at this place
			if(other.place == Environment::get_place()) {
				// If both tasks are spawned locally go depth first
				return depth > other.depth;
			}
			// Prefer local task
			return true;
		}
		else if(other.place == Environment::get_place()) {
			// Only other task was spawned at this place
			// Prefer other task
			return false;
		}
		
		// For non-local tasks go breadth-first
		return depth < other.depth;
	}
private:
	int depth;
};

\end{lstlisting}
\caption{Example strategy for a tree-like algorithm with local
  depth-first execution and breadth-first stealing.}
\label{lst:strategy}
\end{algorithm}

To facilitate locality-aware scheduling, we provide strategy objects
with a way to calculate the memory distance between different places
(not shown in the given example). Using this, strategies can
prioritize tasks for which the data is stored in a nearby place.

\section{Strategy scheduler}
\label{sec:sched}

Our strategy-aware scheduler implements a typical work-stealing
scheduler, but uses a priority data structure (briefly described in
Section~\ref{sec:ds}) per thread instead of a standard work-stealing
deque. Each thread has affinity to a specific CPU core to allow for
locality-centric optimizations, captured in the concept of
\emph{place}.  The abstract machine/memory model used by the scheduler
is a (balanced) tree, where the leaves represent processing units, and
the nodes group processors that share some level in the memory
hierarchy.  The information needed on the concrete machine is gathered
using hwloc~\cite{hwloc10}.
This abstract machine model allows for another locality-specific
optimization, in which tasks are stolen from neighboring processing units
first.
In our scheduler, newly spawned tasks are put into the priority
data structure, and the continuation is executed first.  This
\emph{help-first} scheduling policy~\cite{GuoBarik09} differs from the
\emph{work-first} scheduling policy used in work-stealing systems like
Cilk~\cite{Leiserson10}, where the spawned task is executed
immediately, and the continuation is made available to other threads
to be stolen.  The \emph{help-first} scheduling policy is required for
priority scheduling, as a decision for the execution order of spawned
tasks can only be made if we first generate the tasks and then execute
the task with the highest priority.  The synchronization constructs
used to wait for tasks are \emph{finish} regions as known from
X10~\cite{Charles05}. A finish region ensures that execution continues
after the region only when all tasks spawned inside the region,
including transitively spawned tasks, have completed.  


\subsection{Task storage data structure}
\label{sec:ds}

In work-stealing schedulers each place has its own task-storage
data structure, where new tasks are put in and from which tasks are
executed. Only when its task-storage data structure is empty does a place
access the data structure of another place in an attempt to steal
work. Our data structure must be able to support that each place that
accesses a data structure can prioritize tasks differently. Local
accesses (push and pop accesses by the owner of the data structure)
are common, so the priority ordering for the owner is updated every
time a new task is added using a separate local priority
data structure. Since stealing accesses are rare, the priority
ordering for the place that performs the steal attempt can be
evaluated lazily. This lazily evaluated priority ordering is cached by
the stealer and updated with newly added tasks at the next steal
attempt. We have designed a lock-free implementation of this kind of
data structure, the algorithmic design of which is outside the scope
of this paper.

The evaluation of the priority ordering is performed using a separate
heap-based data structure. This data structure is used both for the
local ordering, as well as for the lazy evaluation of the steal
order. To ensure composability of strategies, the priority
data structure needs to be aware of the different types of strategies
available in the system to first generate an ordering for each type of
strategy, before creating an ordering between the highest-priority
strategies of each type in their parent strategy.

\section{Applications}
\label{sec:applications}

We have selected a number of (kernel) applications that can profit
from scheduling strategies in different ways to
illustrate both advantages and flexibility with their use. We
describe the applications and the customized strategies in this
section; performance results are given in Section~\ref{sec:results}.

\paragraph{Graph Bipartitioning}
\label{sec:bb}

The branch-and-bound paradigm is generally well suited to
parallelization~\cite{CrainicLeCunRoucairol2006}.  Efficient parallel
branch-and-bound implementations rely on a concurrent data structure
for storing unexplored
subproblems~\cite{HerleyPietracaprinaPucci99,KarpZhang93,Sanders95}.
Performance depends crucially on the order in which subproblems are
explored in order to quickly find new feasible solutions to bound
subproblems that do not have to be explored.  We use strategies to
effect the prioritized execution order.

We focus on the well-known, NP-hard \emph{graph bipartitioning
  problem}~\cite{PapadimitriouSteiglitz82} where the vertices of an
undirected, weighted, $n$-node, $m$-edge graph are to be partitioned
into two sets with given sizes with minimum total cut weight. For
bounding (elimination) of subproblems we use a simple, easily
computable lower bound~\cite{Traff91:or} with an
additional improvement for dense graphs. Incrementally updating the
lower bound for each new node subproblem takes $O(n\log
n+m/n)$ (amortized) steps.

\begin{algorithm}
\begin{lstlisting}[mathescape=true,columns=flexible,escapechar=!]
if (sub_problem->lower_bound >= *upper_bound)
  // Bound: a better solution is already known
  return;

// Branch: generate two new subproblems by assigning
// most promising vertex to either subset
SubProblem* sub_problem2 = sub_problem->split();

if (sub_problem->is_solution()) {
  // New feasible solution; update upper bound atomically
  sub_problem->update_solution(upper_bound, solution);
} else if (sub_problem->lower_bound < *upper_bound) {
  spawn_s<BBTask>(
  /* strategy for scheduling */ Strategy(sub_problem, upper_bound),
  /* parameters for task */ sub_problem, upper_bound, solution);
}
// same for other sub_problem
\end{lstlisting}
\caption{Branch-and-bound task with strategies.}
\label{lst:bb}
\end{algorithm}

The C\verb!++! code fragment in Algorithm~\ref{lst:bb} shows task
parallel implementation of the branch-and-bound paradigm.

In our implementation subproblems are represented as tasks. The value
of the currently best known feasible solution is kept in a global
variable that is updated atomically as tasks find better
solutions. When a task is executed it first checks whether the
computed lower bound exceeds the global best known solution (upper
bound). If this is not the case, the problem is split at a chosen
branching vertex and two new subproblems are spawned. At spawn time
subproblems/tasks are assigned a scheduling strategy, which can use
the information known about the subproblems to influence subproblem
exploration order. We use an estimate on the best solution value for
the subproblem which can be computed together with the lower bound in
$O(n)$ steps.  The estimated solution value for each subproblem is
used to prioritize tasks. At each place, the task with the smallest
estimate is executed first.  Since the estimate is mostly decreasing,
this leads to a depth-first execution, where the most promising
branches are executed first.  When a place runs out of tasks, it
steals tasks from another place. We prefer to steal tasks that have a
high \emph{uncertainty}, by which we mean the difference between the
estimated solution value and the lower bound for the subproblem. Such
tasks are likely to generate much work and subproblems
that may lead to a good solution. This reduces further interaction
(stealing) between places.

We also use strategies to
convert spawns into function calls. Many tasks do not generate
much work since they lie on branches that are cut off early. Also, for
many problem instances most of the time is spent not on finding the
best solution, but on verifying that no better solution exists among
the still active subproblems. The overhead for creating and scheduling
those tasks can be significantly reduced by performing call conversion
for tasks that are not expected to generate much work. To do this, the
transitive weight of each task has to be estimated. A rough estimate
on the depth that needs to be explored for a task is given by the
value of the best known solution minus the current lower bound divided
by the average contribution of each node to the best known
solution.
We assign a transitive weight of $2^d-1$ to the task, where $d$
is the estimated depth, under the expectation that the full subtree of
height $d$ has to be generated.

\paragraph{Prefix sum}
\label{sec:prefix}

A typical, parallel, blocked prefix-sums algorithm computes in
parallel the prefix sums for a sequence of distinct blocks, then
computes (recursively) the prefix sums for the sequence block sums,
and finally in parallel for each block adds the previous block prefix
sum to all elements of the block. We can use strategies to reduce the
extra overhead caused by the last step in cases where there is little
parallelism, or where other applications are running at the same
time. The observation is that if for any given block the prefix sums
for the previous block have already been computed, then the previous
block prefix sum can simply be added to the first element of the block
before performing the prefix sums computation. This eliminates the
need for the last step of the above parallel algorithm. This makes
sense in cases where there is little parallelism and many blocks are
processed by the same place; in such cases the performance of the
parallel algorithm can be expected to be on par with a sequential
prefix sums implementation, and not, as would have been the case
without strategies, a factor of two slower. This example illustrates
how strategies can be used to achieve algorithm adaptivity.

With strategies, tasks use a global counter to detect whether the
predecessor block has already been processed: this is the case if the
counter is equal to the block number. The counter is incremented
whenever a block has been processed in order. The strategy ensures
that some place processes blocks in order of increasing block number,
and all other places in decreasing order.

\paragraph{Unbalanced Tree Search}
\label{sec:uts}

The Unbalanced Tree Search (UTS) benchmark by Olivier et
al.~\cite{olivier2007} spawns a large number of small tasks,
corresponding to nodes in a unbalanced search tree, according to a
given distribution.  The decision on how many subtasks to spawn from a
given task is made with the help of a hash of the parent descriptor
and the child index. This makes it possible to get the exact same tree
every time, based on the parameters given to initialize the tree.
This behavior makes the UTS benchmark a candidate for evaluating the
spawn-to-call feature of the strategy scheduler. For our experiments
we use the T5 tree from the UTS benchmark suite. This is a tree with a
geometric distribution and a maximum height of twenty that generates
around four million tasks. Our strategy assigns a high transitive
weight to tasks close to the root and a low weight to tasks closer to
the leaves.  The weight increases exponentially depending on the
distance from the maximum height, but is capped to not grow too large.

\paragraph{Triangle strip generation}


The generation of triangle strips to represent 3D models is a common
optimization for improving rendering performance. Instead of passing
individual triangles to the rendering hardware, adjacent triangles are
combined into strips, where vertices appearing in two adjacent
triangles only have to be transmitted once. This lowers the number of
vertices from $3n$ to $n+2$.  In the optimal case one would need only
one triangle strip to represent the entire model. The optimization problem
is NP-complete and thus best solved using heuristics.

We have used a version of the so called SGI algorithm~\cite{evans96};
the data used is the 3D model Lucy from the Stanford 3D Scanning
Repository\footnote{\url{http://graphics.stanford.edu/data/3Dscanrep}}.
The model consists of around 28 million triangles.  To minimize the
number of single triangle strips a node on the graph is randomly
picked from the set of nodes with the lowest degree.  A strip is then
built by adding neighboring nodes to the strip at both ends. Priority
is given to nodes with a low number of neighbors. When no more nodes
can be added to the strip, a new node is randomly picked and a new
strip is started. This is repeated until all nodes are part of a
strip.

With this benchmark we aim to show that strategies can lead to both
qualitatively better results and performance improvements.  To improve
the result and generate fewer and longer triangle strips, strategies
prioritize picking of nodes (tasks) with a low number of neighbors,

The benchmark uses two types of tasks. The first is the StartTask,
which at spawn-time is assigned a pointer to a possible node to start
a triangle strip from. The strategy for this type of task stores the
number of neighboring nodes that are not part of a strip, and it uses
that to prioritize tasks. Generating a strip is a relatively quick
operation, so it is suitable for spawn-to-call transformation and is
thus given a low transitive weight. Spawning a StartTask for every
node in the graph would be wasteful, as many will be part of other
triangle strips and thus not eligible to start a new strip
from. Instead we provide a second type of task, the SpawnTask, to
gradually spawn new StartTasks when needed and only for eligible start
nodes. This task spawns new StartTasks for a given interval of
nodes. The strategy used has a transitive weight which is the same as
the number of tasks that it will spawn. It does not allow the task to
be transformed into a call.  The two strategies are composed by a
common parent strategy that gives priority to SpawnTasks when stealing
and to StartTasks when working locally.

\paragraph{Single-source shortest path}
\label{sec:sssp}

Single-source shortest path again shows how algorithms that require
prioritization of work can be parallelized in a simple way with
strategies. We use an obvious parallelization of Dijkstra's
algorithm. Tasks update distance labels and the role of the priority
queue is taken over by the task scheduler; this same, straightforward
parallelization is also used by Lenharth et
al.~\cite{pingali2011}. Note that although this type of
parallelization may work well in average, it cannot guarantee any
speed-up (some places may simply do superfluous work that a sequential
execution would not), although it should never perform worse than a
sequential algorithm using the same priority data structure.

The strategy for the owning thread is to explore
the most promising path first. This is the task with the smallest
distance value. Stealing all promising tasks
might be a bad idea, since then the original owner would end up with
only less promising tasks to explore.
Instead, we steal random tasks. To effect this, a random number is
created for each instance of the strategy, and strategies are ordered
by this random number for stealing accesses.

\paragraph{Quicksort}
\label{sec:qs}

Even simple, standard-example kernels can profit from strategies as we
show with a standard, task-parallel Quicksort algorithm (with a
sequential partitioning algorithm).  The best cache behavior is
expected if locally spawned tasks are executed depth-first, and the
shorter subsequence is processed first. When stealing tasks, the
largest subsequences should be stolen first to reduce interference.
For the standard LIFO/FIFO strategy typically used by work-stealing
schedulers, it is easy to see that most of those criteria are already
fulfilled, except for choosing the smaller subsequence when going
depth-first.  Therefore, only small gains can be expected from
choosing the smaller subsequence.  More performance gains can be
expected by converting task spawns to function calls, when enough
tasks are present in the task queue, and by choosing a better number
of tasks to steal. We achieve this by configuring a transitive weight
for each Quicksort task and by enabling call conversion. The expected
running time of Quicksort for a sequence of length $n$ is $O(n\log
n)$, so the transitive weight is set to $n'\log n'$, where $n'=n/b$
for some block size $b$ (a tuning parameter; as a rule of thumb $b$ is
chosen such that the transitive weight for the smallest task worth
parallelizing is~$1$).

\section{Experimental results}
\label{sec:results}

We have implemented and benchmarked the applications presented in
Section~\ref{sec:applications} with strategies as discussed.  An
implementation that uses a \emph{LIFO/FIFO strategy}
(\emph{last-in-first-out} order for local tasks,
\emph{first-in-first-out} order for tasks being stolen with behavior
similar to the Arora et al.\ work-stealing
deques~\cite{AroraBlumofePlaxton01}) is used as a baseline for
comparison.  To evaluate the overhead of the strategy enhanced
scheduling framework we also compared the results to implementations
executed with a standard work-stealing scheduler.  To ensure a fair
comparison, this work-stealing scheduler is implemented with the same
techniques and optimizations as the strategy scheduler, but without
support for strategies, and with a wait-free implementation of the
Arora et al.\ deque~\cite{AroraBlumofePlaxton01}. We benchmarked this
scheduler against Cilk\verb!++!~\cite{Leiserson10} and Intel Threading
Building Blocks~\cite{KukanovVoss07} to validate that performance is
comparable to other work-stealing systems, which it is (not shown in
this paper).

\paragraph{System and Settings}

All applications have been executed on a system with four 12-core AMD
Opteron 6168 processors, for a total of 48 cores. It has 128\,GB of
memory, which is sufficient for all benchmarks to be processed in
memory.  The operating system is Linux (Debian 6.0), and the
schedulers use pthreads as the threading layer. All applications have
been compiled using \texttt{g++ 4.7}. The system is a NUMA system, and
often the limited total memory bandwidth makes it hard to achieve
speed-up beyond 12 cores, as experience with other memory bound
(OpenMP) applications has shown.

Experiments were repeated 10 times and the average execution time is
presented. For randomly generated problem instances, 10 different
random seeds were used, but each test used the same 10 seeds and
therefore the exact same problem instances to make scalability results
comparable.

\paragraph{Graph Bipartitioning}
\label{sec:results_bb}

The graph bipartitioning application described in Section~\ref{sec:bb}
has been run with weighted and unweighted random graphs ($G_{n,p}$)
with different sizes and densities. We present results for weighted
graphs with 35 nodes, average density of 90\% and randomly chosen
integer weights in the range $[1,1000]$, and unweighted graphs with 39
nodes and 50\% density. Other inputs showed similar behavior.

\begin{figure}
\centering
\subfigure[Execution time]{
  \includegraphics[width=0.45\textwidth]{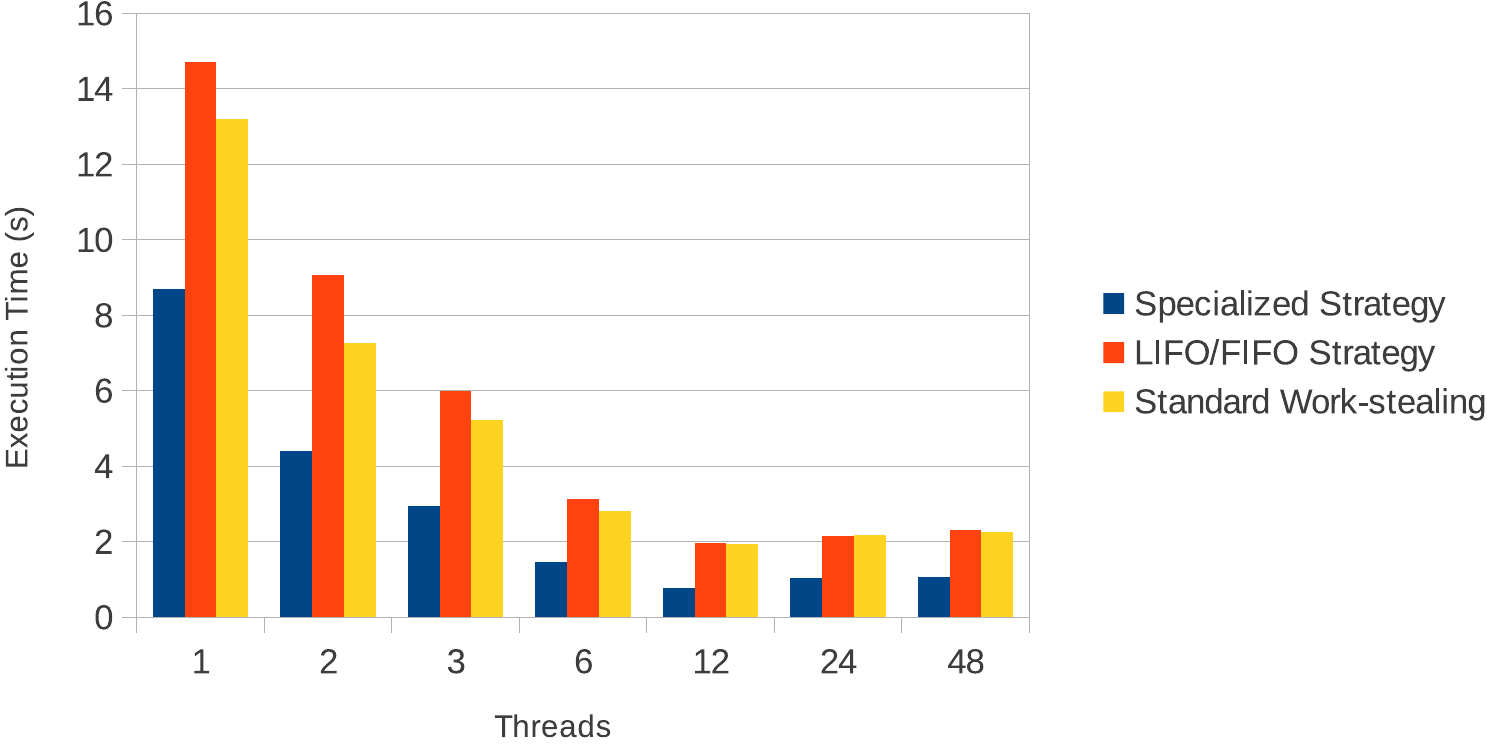}
  \label{fig:bb_saturn_unweighted}
}
\subfigure[Time until optimum found]{
  \includegraphics[width=0.45\textwidth]{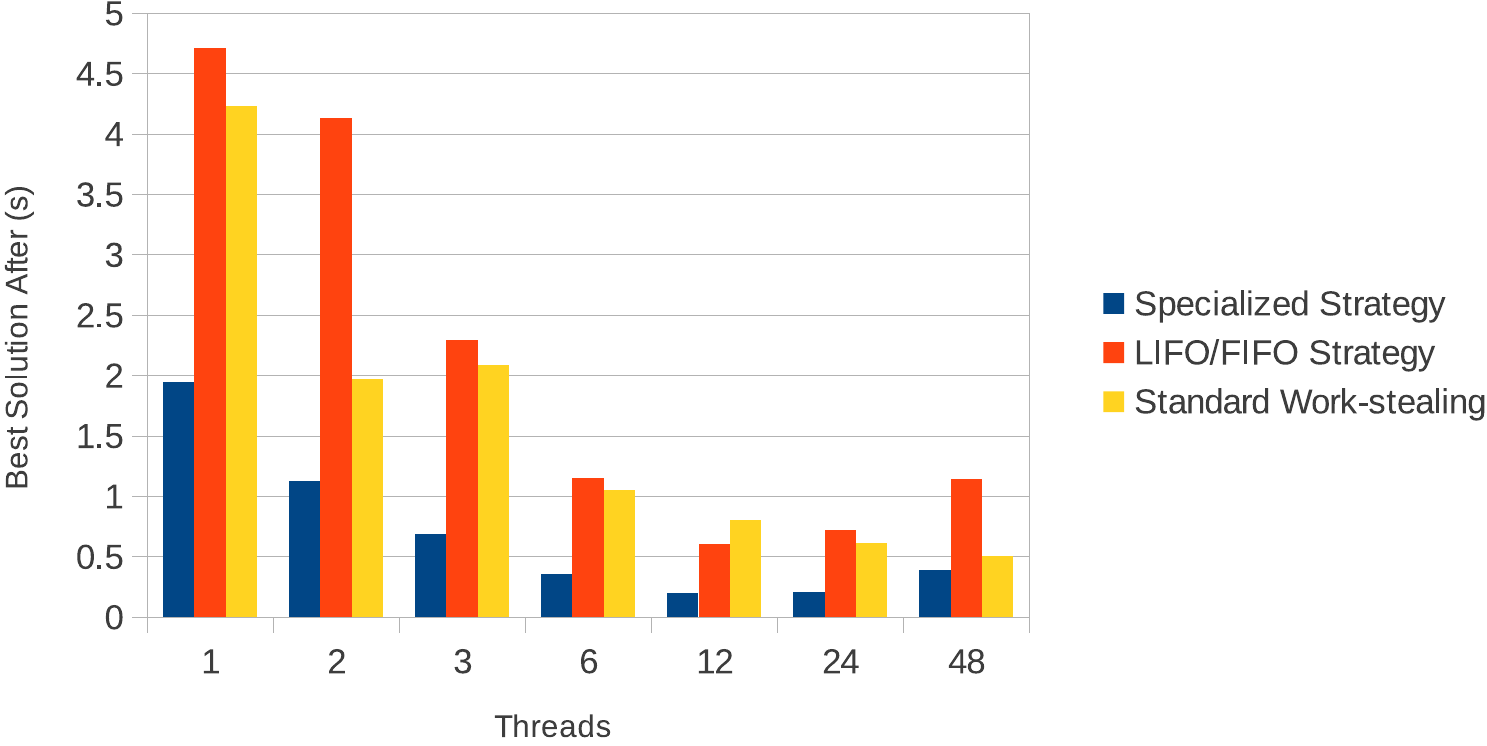}
  \label{fig:bb_saturn_unweighted_bound}
}
\caption{Unweighted graph partitioning. Problem size $n=39$, density: 50\%.}
\end{figure}

The results for the unweighted graphs are shown in
Figure~\ref{fig:bb_saturn_unweighted}.  Independently of the number of
threads in the system, strategies improve the execution time by nearly
a factor of two. Without a specialized strategy, the strategy
scheduler is, as expected, slightly slower than standard
work-stealing.  The overhead for the strategy scheduler becomes less
prominent with increasing number of threads.

The performance advantages from strategies come from two
factors. First, a good strategy leads to an optimal solution being
found sooner, which reduces the total amount of work
necessary. Second, converting task spawns to function calls reduces
overhead, and stealing half the (estimated) work instead of half the
tasks leads to better load balance and less steal attempts. To
illustrate the influence of both factors, we also recorded when the
optimal solution value was found, which corresponds to the last time
the currently best solution was updated.

The results of this measurement are shown in
Figure~\ref{fig:bb_saturn_unweighted_bound}. This metric is less
robust, but it gives insight into how the prioritization of tasks
influences the search for the optimal value. As can be seen, the
difference between the specialized strategy and standard work-stealing
becomes significant, and in some cases reaches a factor of
three. After the optimal solution is found, all implementation
variants generate the same amount of work for a specific branch, since
the same algorithm for pruning branches is used. Nonetheless, there is
still a significant difference between execution times in
Figure~\ref{fig:bb_saturn_unweighted}, when we subtract the time the
optimal value was found. This performance difference is due to the
other optimizations accomplished with strategies, namely call
conversion and stealing half of the work.

\begin{figure}
\centering
\subfigure[Execution time]{
  \includegraphics[width=0.45\textwidth]{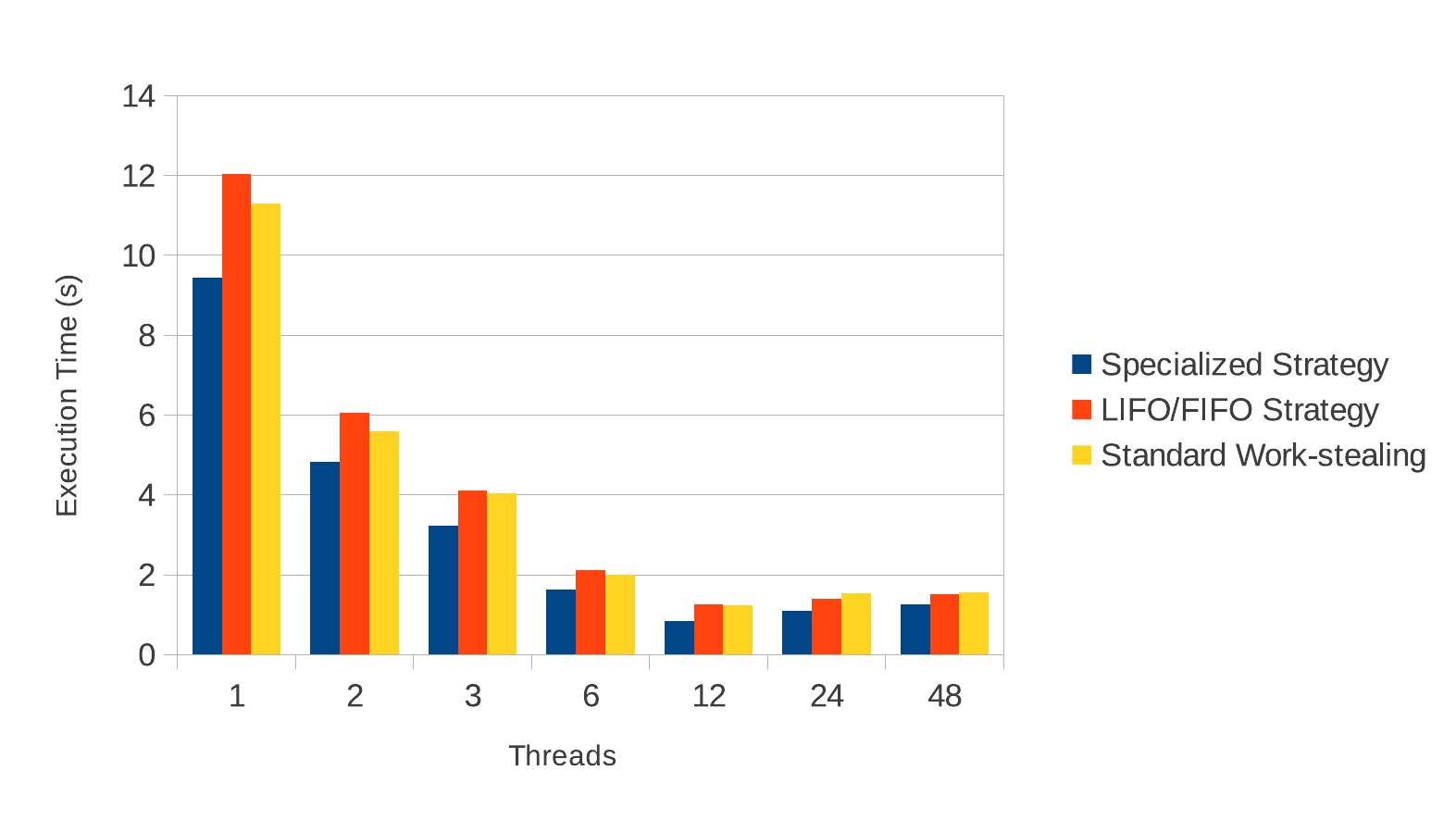}
}
\label{fig:bb_saturn_weighted}
\subfigure[Time until optimum found]{
  \includegraphics[width=0.45\textwidth]{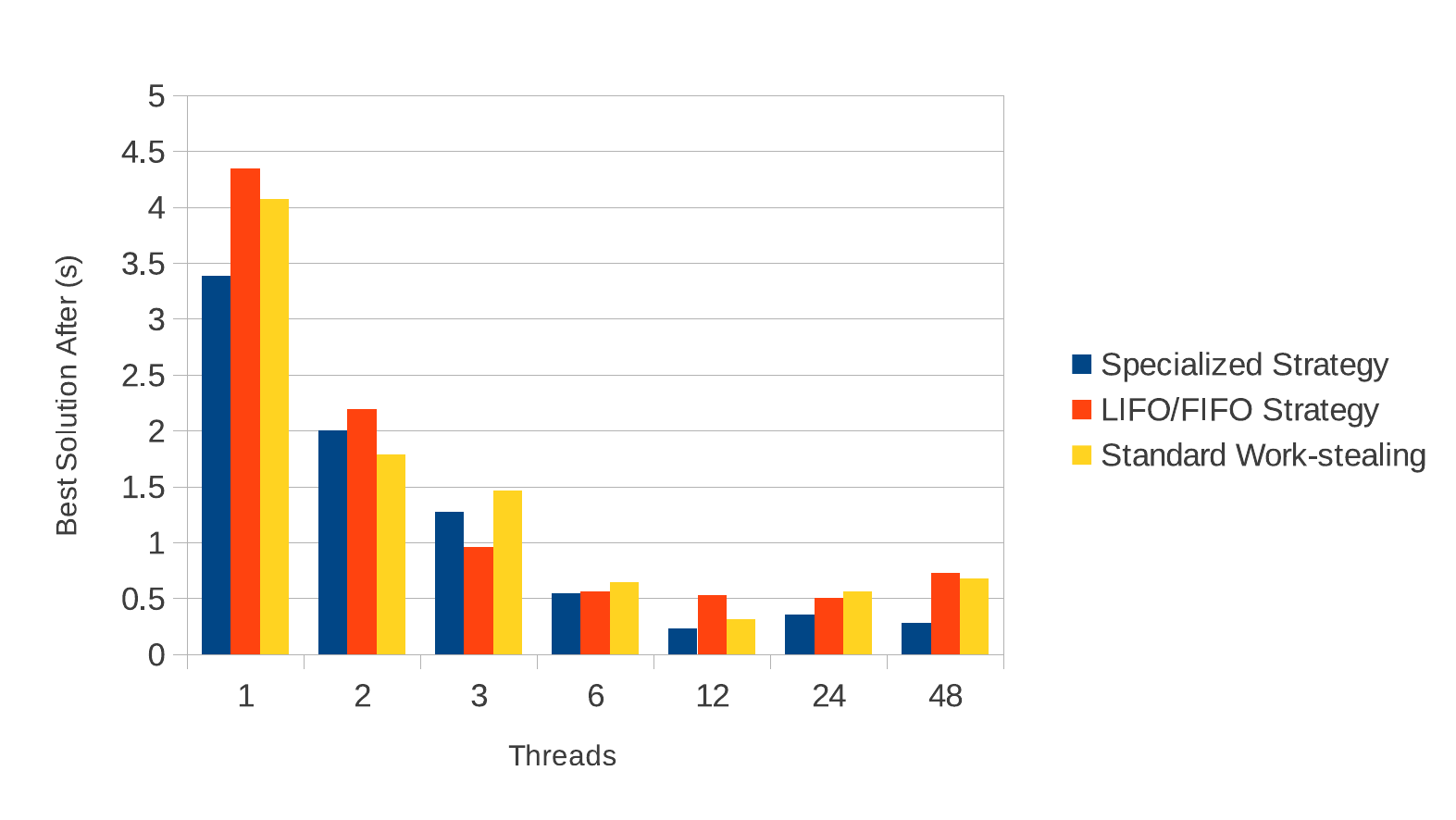}
}
\caption{Weighted graphs. Problem size $n=35$, density: 90\%.}
\label{fig:bb_saturn_weighted_bound}
\end{figure}

Figure~\ref{fig:bb_saturn_weighted} gives the corresponding
measurements for the weighted graph instances.  Since our simple
estimates for best cut value and amount of work generated by a
subproblem become less precise for weighted graphs, we expect smaller
gains by using strategies than in the unweighted case.  Nonetheless,
graph partitioning with strategies still outperforms the standard
work-stealing scheduler. The optimal solution is again found
faster with the strategy than without it, as is shown in
Figure~\ref{fig:bb_saturn_weighted_bound}. However, the difference is
smaller, and in some cases the work-stealing scheduler was more
lucky in finding an optimum fast.

\paragraph{Prefix sum}

\begin{figure}
\centering
\subfigure[1 array]{
  \includegraphics[width=0.45\textwidth]{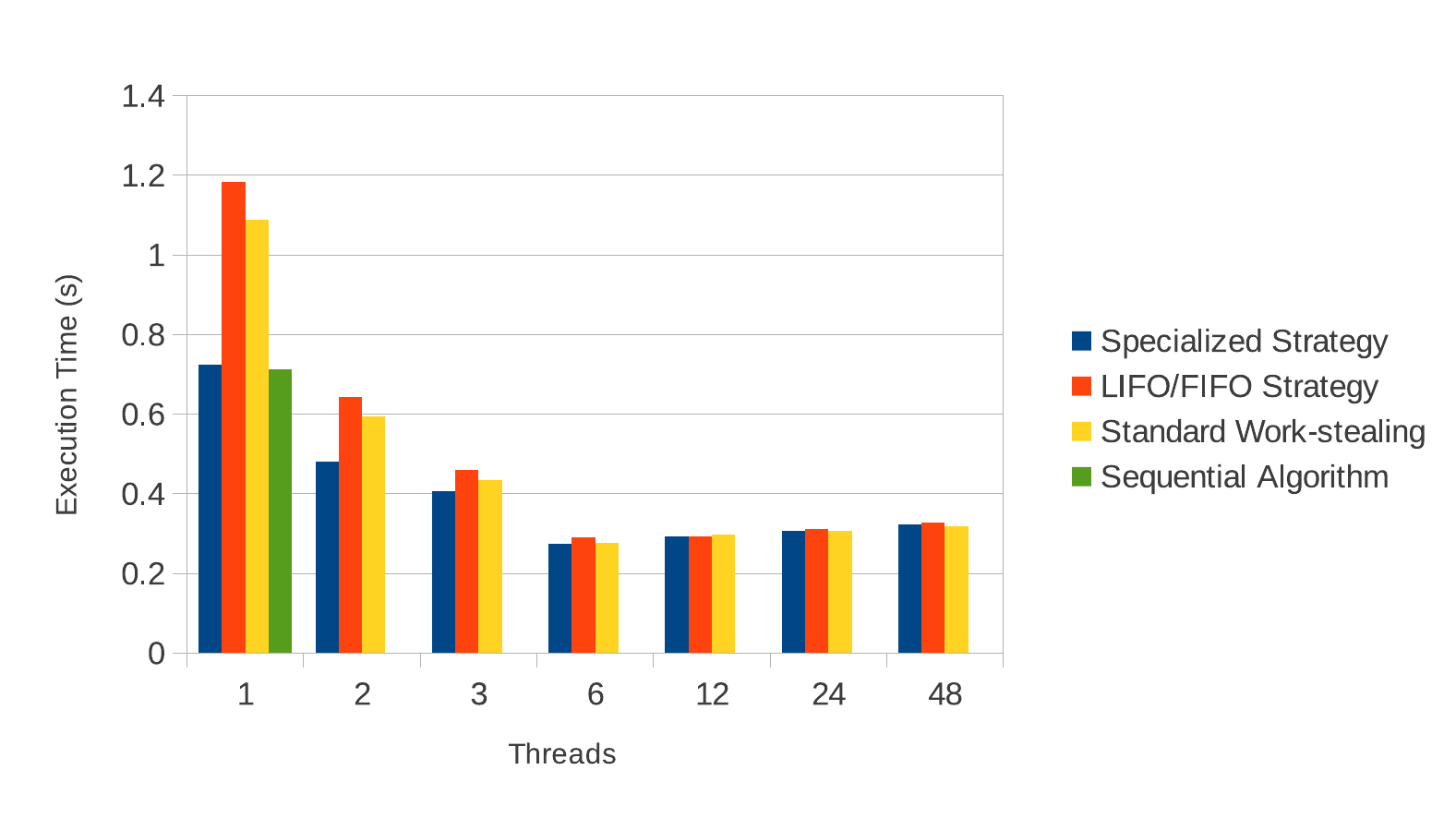}
  \label{fig:scan_saturn_1}
}
\subfigure[12 arrays]{
  \includegraphics[width=0.45\textwidth]{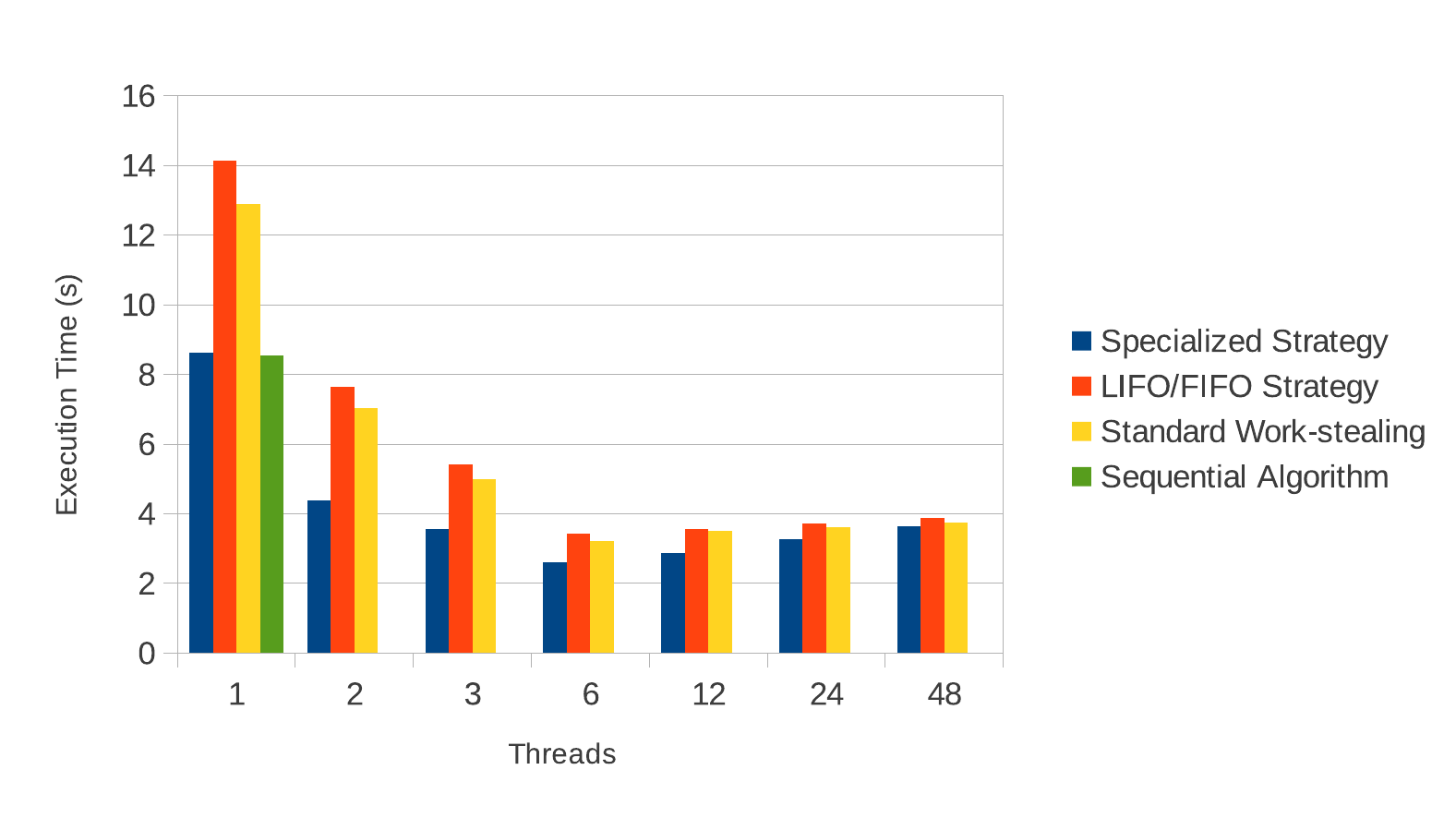}
  \label{fig:scan_saturn_12}
}
\caption{Prefix sum for arrays with $2\times10^8$ elements each.}
\end{figure}

The prefix sum benchmark has been run on an array of $2\times 10^8$
integers with a block size of 4096 elements for the parallel
algorithm.  The execution times are shown in
Figure~\ref{fig:scan_saturn_1}.  In this example strategies are used
to inflict the parallel algorithmic overhead only when enough threads
are available. Since the specialized strategy implementation matches
the performance of the sequential prefix sum algorithm when run on a
single thread, this is obviously the case. For larger numbers of
threads the advantages of the strategy are diminishing, and for 12
threads there is already no noticeable difference between a standard
work-stealing implementation and the specialized strategy. Thus, we
see that strategies achieve the effect of making the algorithm adapt
to the number of available threads at any given time of its execution.

To highlight the adaptive behavior, we measured the performance of 12
concurrent runs of the prefix sum computation within the same single
scheduler (this could be part of an application that is heavy on
prefix sums computations). The results are shown
Figure~\ref{fig:scan_saturn_12}. When using only one thread, the
results are similar, but for larger numbers of threads the
implementation with strategies yields better performance than running
the 12 simultaneous prefix-sums computations with standard
work-stealing.

\begin{figure}
\centering
\begin{minipage}[t]{0.45\textwidth}
\includegraphics[width=\linewidth]{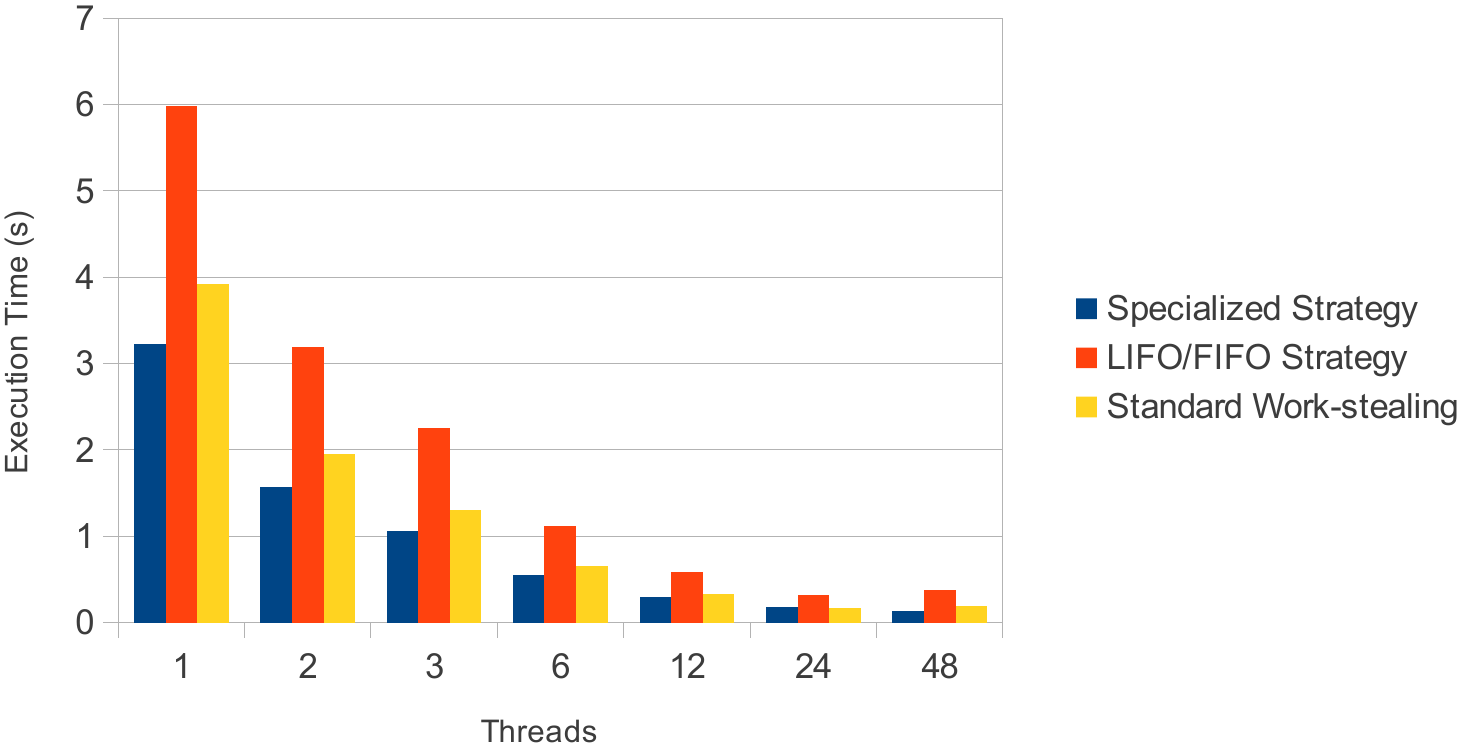}
\caption{Universal tree search for graph T5 with a geometric distribution and a max depth 20.}
\label{fig:uts_result}
\end{minipage}
\hspace{0.05\textwidth}
\begin{minipage}[t]{0.45\textwidth}
\includegraphics[width=\linewidth]{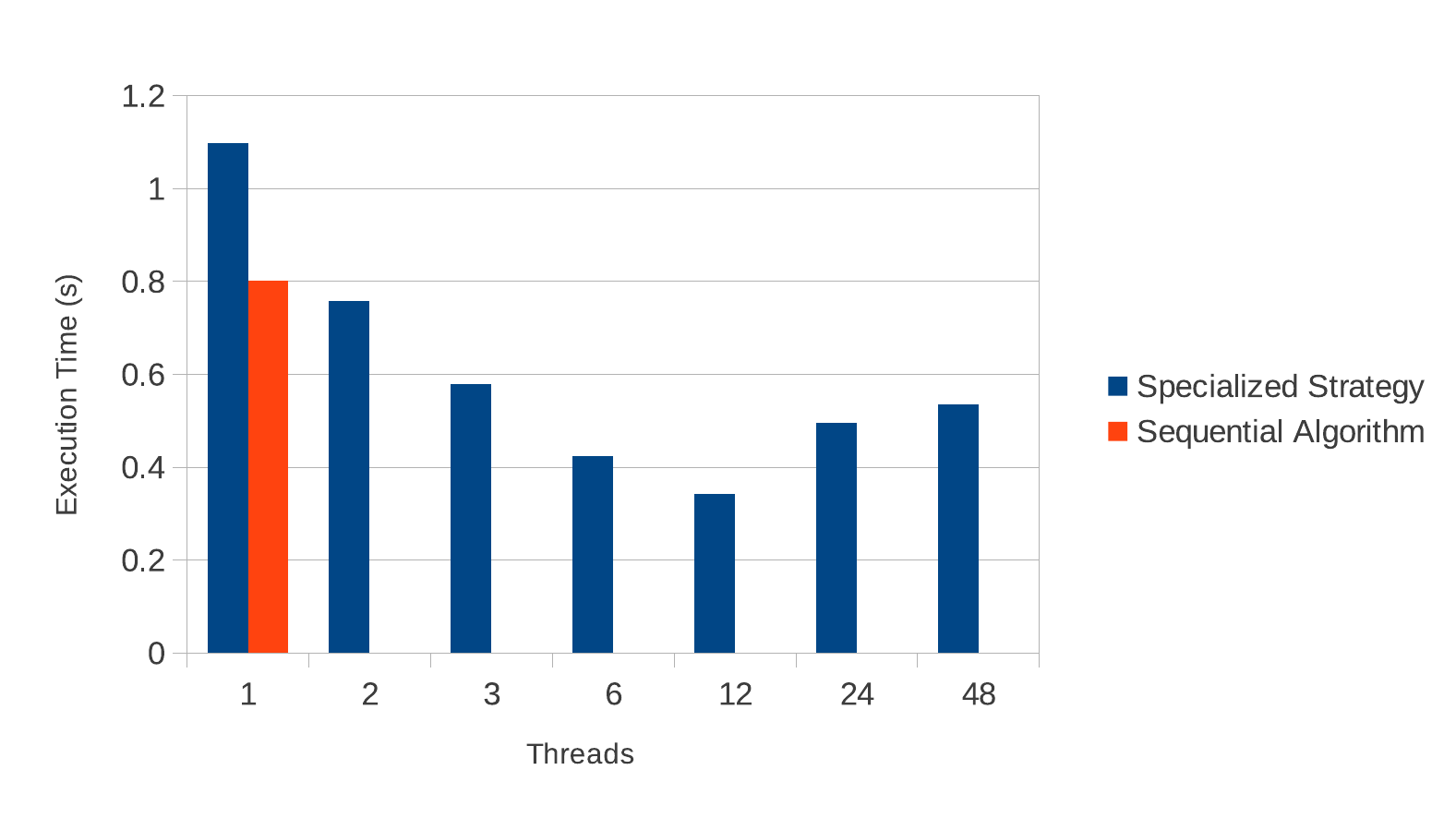}
\caption{Single-source shortest path calculation on a weighted graph. Problem size $n=15000$, density 50\%, integer weights in $[1,1000]$.}
\label{fig:sssp_saturn_sparse_weighted}
\end{minipage}
\end{figure}

\paragraph{Unbalanced Tree Search}

The UTS benchmark creates a large number of small tasks in a short
time-frame. This adds an unnecessary overhead to the task storage
data structure, as can be seen for the LIFO/FIFO-strategy in
Figure~\ref{fig:uts_result}. By allowing smaller tasks, far down in
the tree, to be executed immediately, we lower the churn on the task
storage, which greatly improves the performance. The scheduler decides
on when to convert a spawn to a call by using its knowledge of the
number of tasks currently in the task-storage and the transitive
weight of the task given by its associated strategy. The spawn-to-call
optimization causes our scheduler to outperform the standard
work-stealing scheduler for this benchmark.

\paragraph{Triangle Strip Generation}

\begin{figure}
\centering
\subfigure[Execution time]{
	\includegraphics[width=0.45\textwidth]{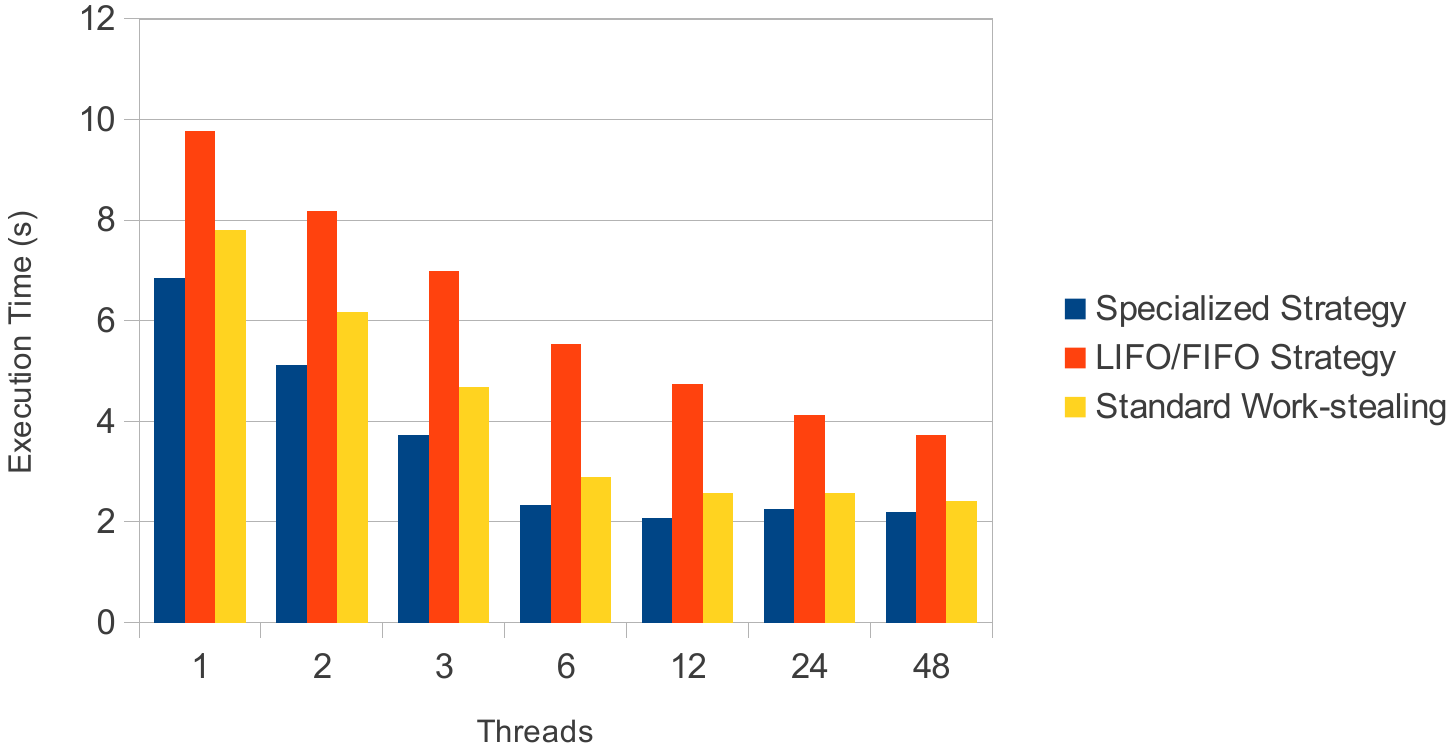}
\label{fig:tristrip_saturn_lucy}
}
\subfigure[Total number of triangle strips generated (lower is better)]{
	\includegraphics[width=0.45\textwidth]{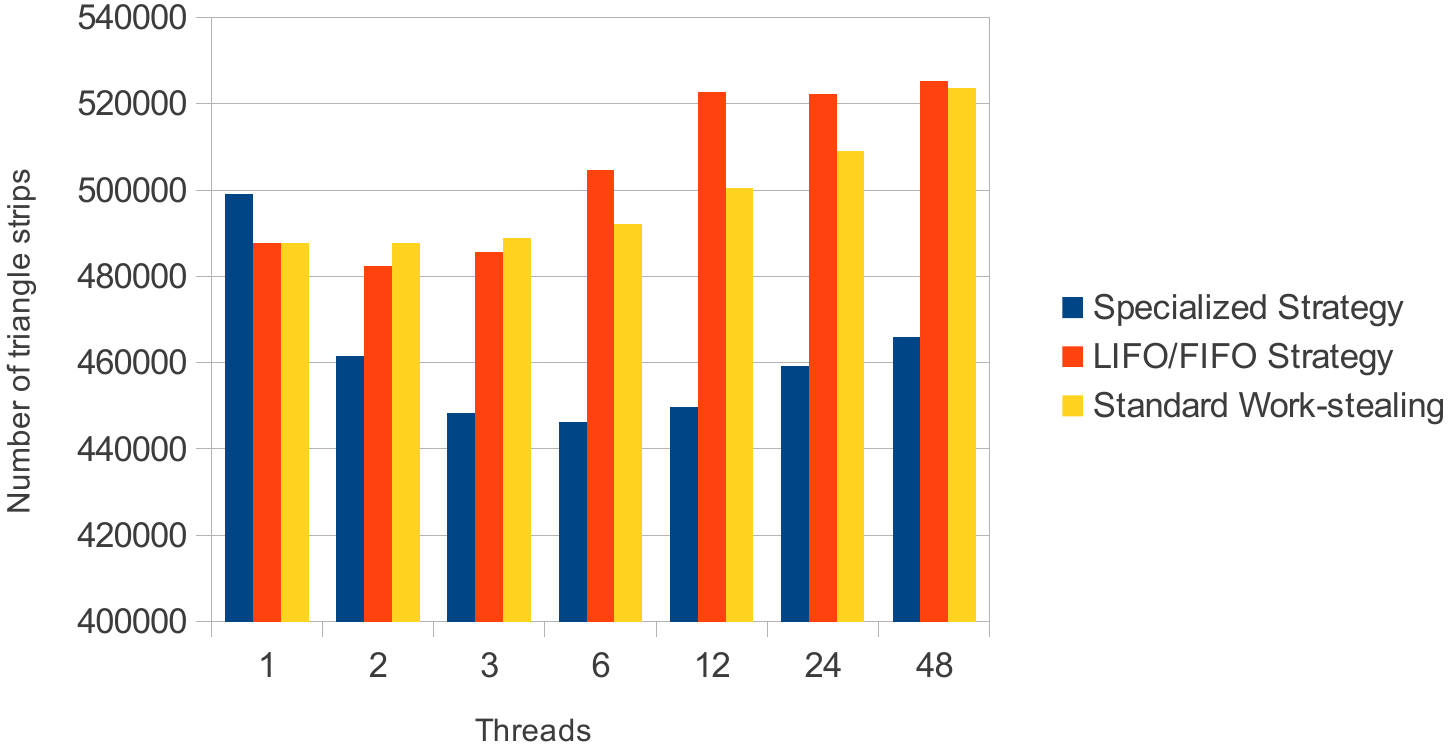}
\label{fig:tristrip_saturn_lucy_count}
}
\caption{Triangle strip generation for the Lucy model with 28 million triangles.}
\end{figure}

The strategy used for triangle strip generation is mainly meant to
improve the qualitative result of the algorithm. However, as the
benchmark is generating a large number of relatively small tasks, we
also get a performance improvement from spawn-to-call conversion.
Figure~\ref{fig:tristrip_saturn_lucy} shows that strategies
perform better than the work-stealing with a fixed order.

Figure~\ref{fig:tristrip_saturn_lucy_count} shows the number of
triangle strips generated (lower is better). The heuristics used are
not guaranteed to give a better result, and there is a great deal of
randomness involved in the algorithm. Despite this we obtain a better
result in a shorter amount of time than compared to the other
schedulers.

\paragraph{Single-source shortest path}

For the single-source shortest path benchmark running the algorithm
with standard work-stealing makes no sense, since it could well take
exponential time when distance updates are performed in some fixed
LIFO order.  We therefore compare to a sequential implementation of
Dijkstra's algorithm with a worst-case efficient priority queue data
structure. The results for weighted graphs with 15000 nodes can be
seen in Figure~\ref{fig:sssp_saturn_sparse_weighted}. Dijkstra's
algorithm slightly outperforms the strategy scheduler in the
sequential setting. The parallel algorithm achieves some scalability,
with a speedup of $~3.3$ over the sequential algorithm on 12 threads.

\paragraph{Quicksort}

\begin{figure}
\centering
\begin{minipage}[t]{0.45\textwidth}
\includegraphics[width=\linewidth]{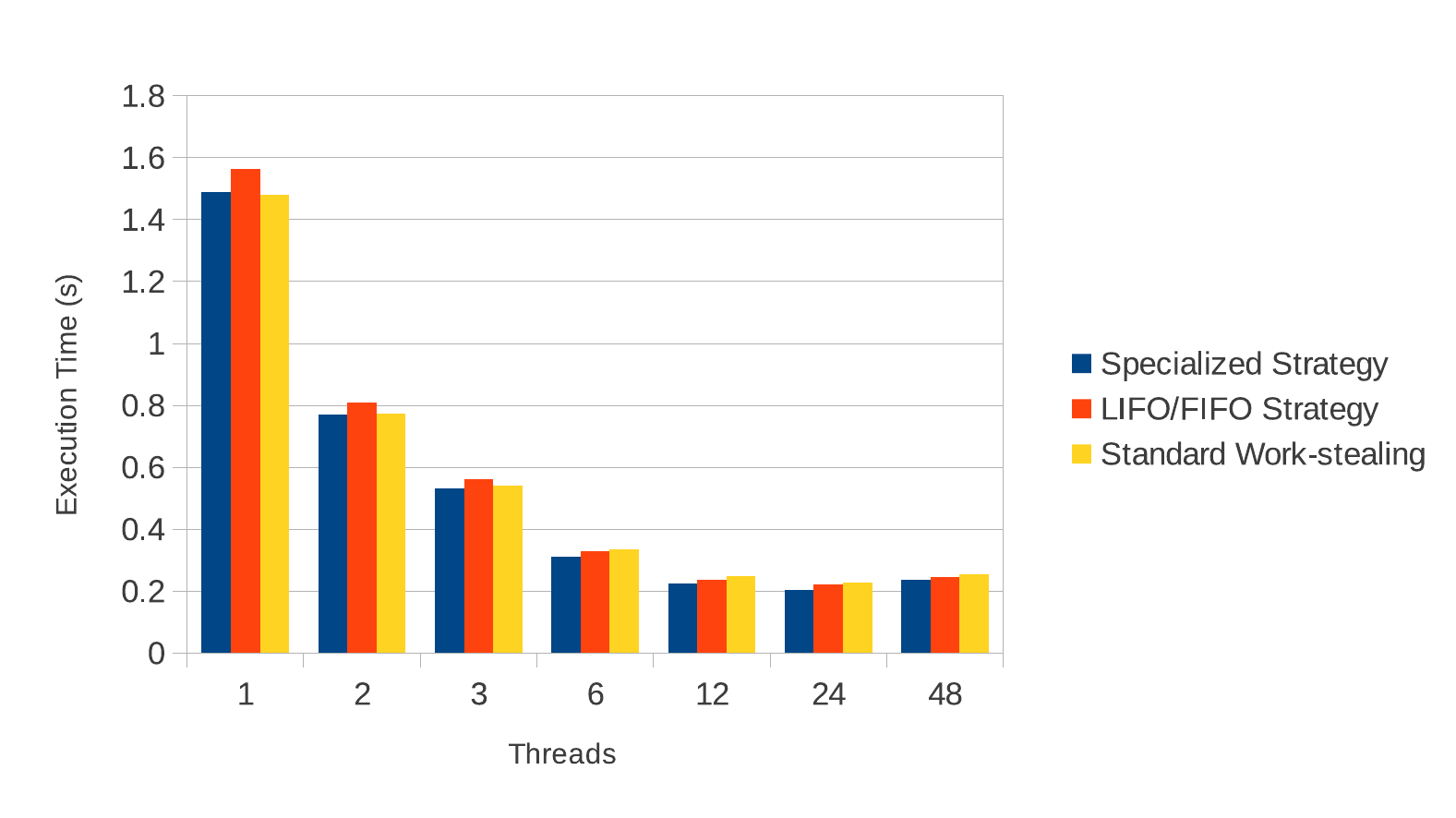}
\caption{Quicksort on a sequence of 10 million elements.}
\label{fig:sorting_saturn}
\end{minipage}
\hspace{0.05\textwidth}
\begin{minipage}[t]{0.45\textwidth}
\includegraphics[width=\linewidth]{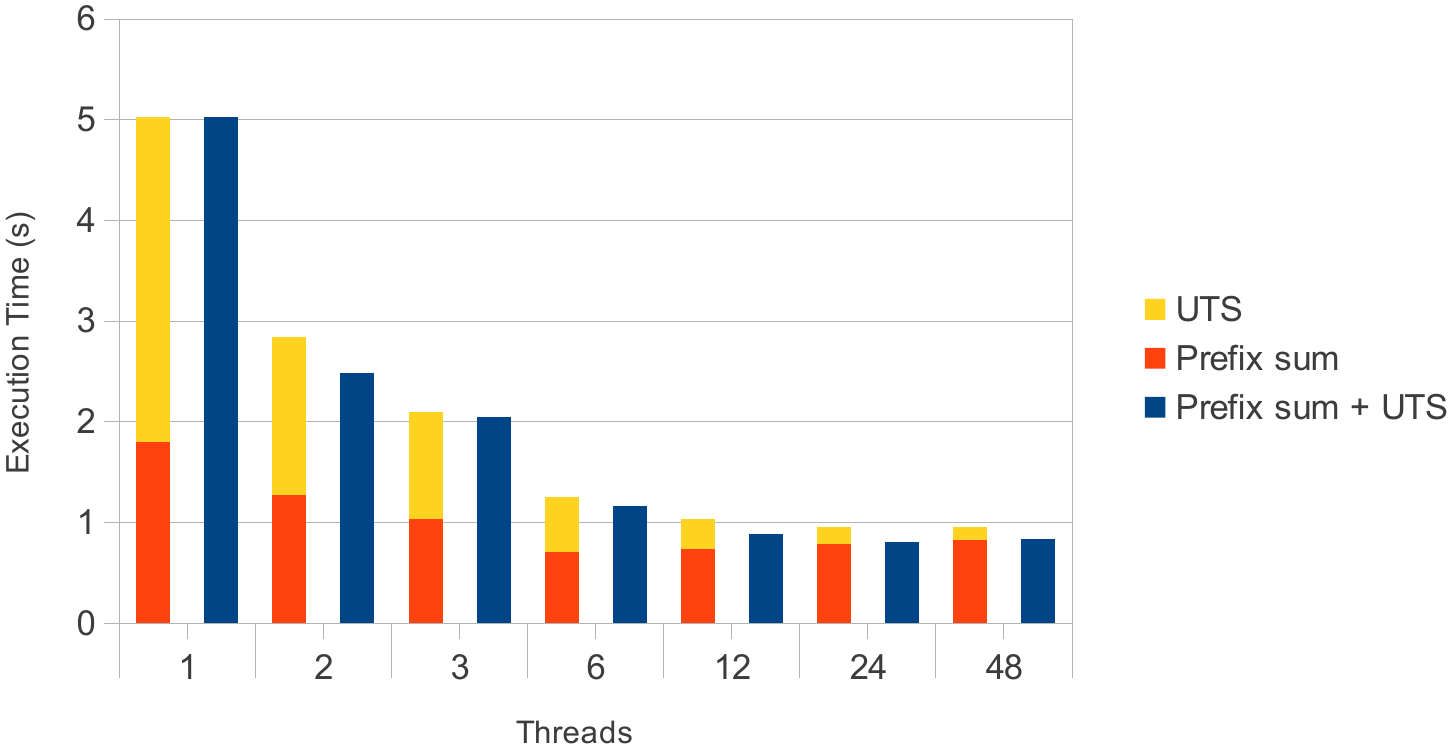}
\caption{Prefix sum for a list with $5\times10^8$ elements together with the UTS benchmark for the T5 tree. All measurements were done using specialized strategies.}
\label{fig:prefix_and_uts_benchmark}
\end{minipage}
\end{figure}

For Quicksort, which fits with the standard work-stealing execution
order, we do not expect much from an execution with a specialized
strategy. We chose Quicksort as a benchmark to see whether some
performance advantage can still be achieved with strategies, and how
high the overhead of the strategy scheduler is. The implementation
uses a cut-off for subsequences shorter than 256 elements, which are
quicksorted sequentially. Figure~\ref{fig:sorting_saturn} shows the
results of those measurements. For the 1-thread run performance with
strategies is similar to standard work-stealing, since the overhead
for the strategy scheduler is roughly the same as the gains due to the
optimizations with strategies. With more threads the overhead
diminishes, and a slightly better performance is observed.

The specialized strategy is able to match the performance of the basic
scheduler for low numbers of threads and can slightly outperform it
for higher numbers. Given that Quicksort is a well-behaving algorithm
for work-stealing this shows that even in such applications a
potential for strategies exists.

\paragraph{Composition of Prefix Sum and UTS}
\label{sec:scan_uts}

To finally demonstrate the performance composability of strategies, we
combined the prefix sum and the UTS benchmark into a single
application. The UTS part uses the same tree as when executed by
itself, whereas the prefix sum was given a slightly longer sequence to
work on, to balance the total amount of work of the two kernels.
The results for the composite benchmark are compared to the
running times of each of the benchmarks on their own using their
specialized strategies. The results in
Figure~\ref{fig:prefix_and_uts_benchmark} show that the performance of
the composed benchmark is better than the sum of its parts for two or
more threads. For a single thread, similar performance is achieved.

\section{Conclusion}

We introduced dynamic scheduling strategies for work-stealing
schedulers to enable application dependent, per-task scheduling
decisions, like changing the execution and stealing order of tasks, as
well as merging tasks at run-time. These decisions can be used to
reduce scheduling overhead, as well as make the execution more
efficient and adaptive.  In contrast to global scheduling policies,
our strategies can be selected on the level of individual tasks. This
poses the unique challenge of making strategies composable, which is
especially difficult for the prioritization of tasks, without
increasing the scheduler overhead too much. However, the focus of this
paper was solely on presenting the idea of scheduling strategies and
giving a glimpse of what can be achieved using this mechanism.

We discussed a variety of applications that profit from such
strategies and presented experimental results to show the expected
gains. The paper gave but a brief, high-level overview of the
scheduling system and the corresponding data structure; details will
be exposed in an accompanying publication.

In future work we plan to explore additional aspects of strategies
that can be used for additional optimization, but also to support more
general models of parallelism like mixed-mode parallelism, where tasks
in themselves can be parallel.  We will also explore further
applications, with the focus on applications that can profit from
different types of strategies in a single execution to investigate the
composability aspect further. Also, there is still room for decreasing
the scheduler overhead by additional optimizations to the
data structure. Finally, we plan to extend our work to heterogeneous
systems, since we believe that strategies can be highly profitable to
aid scheduling decisions in this context.

\bibliographystyle{abbrv}
\bibliography{strategies}

\begin{thebibliography}{10}

\bibitem{AcarBlellochBlumofe02}
U.~A. Acar, G.~E. Blelloch, and R.~D. Blumofe.
\newblock The data locality of work stealing.
\newblock {\em {T}heory of {C}omputing {S}ystems}, 35(3):321--347, 2002.

\bibitem{AroraBlumofePlaxton01}
N.~S. Arora, R.~D. Blumofe, and C.~G. Plaxton.
\newblock Thread scheduling for multiprogrammed multiprocessors.
\newblock {\em {T}heory of {C}omputing {S}ystems}, 34(2):115--144, 2001.

\bibitem{Berenbrink01}
P.~Berenbrink, T.~Friedetzky, and L.~A. Goldberg.
\newblock The natural work-stealing algorithm is stable.
\newblock In {\em 42nd {IEEE} Symposium on Foundations of Computer Science
  {(FOCS)}}, pages 178--187, 2001.

\bibitem{BlumofeJoergKuszmaulLeisersonRandallZhou96}
R.~D. Blumofe, C.~F. Joerg, B.~C. Kuszmaul, C.~E. Leiserson, K.~H. Randall, and
  Y.~Zhou.
\newblock Cilk: An efficient multithreaded runtime system.
\newblock {\em {J}ournal of {P}arallel and {D}istributed {C}omputing},
  37(1):55--69, 1996.

\bibitem{BlumofeLeiserson99}
R.~D. Blumofe and C.~E. Leiserson.
\newblock Scheduling multithreaded computations by work stealing.
\newblock {\em Journal of the {ACM}}, 46(5):720--748, 1999.

\bibitem{hwloc10}
F.~Broquedis, J.~Clet-Ortega, S.~Moreaud, N.~Furmento, B.~Goglin, G.~Mercier,
  S.~Thibault, and R.~Namyst.
\newblock hwloc: A generic framework for managing hardware affinities in {HPC}
  applications.
\newblock In {\em 18th Euromicro Conference on Parallel, Distributed and
  Network-based Processing {(PDP)}}, pages 180--186, 2010.

\bibitem{Charles05}
P.~Charles, C.~Grothoff, V.~Saraswat, C.~Donawa, A.~Kielstra, K.~Ebcioglu,
  C.~von Praun, and V.~Sarkar.
\newblock X10: an object-oriented approach to non-uniform cluster computing.
\newblock In {\em 20th {ACM} {SIGPLAN} Conference on Object-Oriented
  Programming, Systems, Languages, and Applications {(OOPSLA)}}, pages
  519--538, 2005.

\bibitem{Traff91:or}
J.~Clausen and J.~L. Tr{\"{a}}ff.
\newblock Implementation of parallel branch-and-bound algorithms -- experiences
  with the graph partitioning problem.
\newblock {\em Annals of Operations Research}, 33:331--349, 1991.

\bibitem{ColeRamachandran10}
R.~Cole and V.~Ramachandran.
\newblock Resource oblivious sorting on multicores.
\newblock In {\em Automata, Languages and Programming, 37th International
  Colloquium ({ICALP}) Proceedings, Part I}, volume 6198 of {\em {L}ecture
  {N}otes in {C}omputer {S}cience}, pages 226--237, 2010.

\bibitem{CrainicLeCunRoucairol2006}
T.~G. Crainic, B.~L. Cun, and C.~Roucairol.
\newblock Parallel branch-and-bound algorithms.
\newblock In E.-G. Talbi, editor, {\em Parallel Combinatorial Optimization},
  pages 1--28. Wiley, 2006.

\bibitem{evans96}
F.~Evans, S.~Skiena, and A.~Varshney.
\newblock Optimizing triangle strips for fast rendering.
\newblock In {\em 7th IEEE Conference on Visualization}, pages 319--326, 1996.

\bibitem{GuoBarik09}
Y.~Guo, R.~Barik, R.~Raman, and V.~Sarkar.
\newblock Work-first and help-first scheduling policies for async-finish task
  parallelism.
\newblock In {\em 23rd {IEEE} International Parallel and Distributed Processing
  Symposium {(IPDPS)}}, pages 1--12, 2009.

\bibitem{GuoZhaoCaveSarkar10}
Y.~Guo, J.~Zhao, V.~Cav{\'e}, and V.~Sarkar.
\newblock {SLAW}: A scalable locality-aware adaptive work-stealing scheduler.
\newblock In {\em 24th {IEEE} International Parallel and Distributed Processing
  Symposium {(IPDPS)}}, pages 1--12, 2010.

\bibitem{HerleyPietracaprinaPucci99}
K.~T. Herley, A.~Pietracaprina, and G.~Pucci.
\newblock Fast deterministic parallel branch-and-bound.
\newblock {\em Parallel Processing Letters}, 9(3):325--333, 1999.

\bibitem{KarpZhang93}
R.~M. Karp and Y.~Zhang.
\newblock Randomized parallel algorithms for backtrack search and
  branch-and-bound computation.
\newblock {\em Journal of the {ACM}}, 40(3):765--789, 1993.

\bibitem{KukanovVoss07}
A.~Kukanov and M.~J. Voss.
\newblock The foundations for scalable multi-core software in {Intel Threading
  Building Blocks}.
\newblock {\em Intel Technology Journal}, 11(4), 2007.

\bibitem{Leiserson10}
C.~E. Leiserson.
\newblock The {Cilk++} concurrency platform.
\newblock {\em The Journal of Supercomputing}, 51(3):244--257, 2010.

\bibitem{pingali2011}
A.~Lenharth, D.~Nguyen, and K.~Pingali.
\newblock Priority queues are not good concurrent priority schedulers.
\newblock Technical Report TR-11-39, Department of Computer Science, The
  University of Texas at Austin, 2011.

\bibitem{olivier2007}
S.~Olivier, J.~Huan, J.~Liu, J.~Prins, J.~Dinan, P.~Sadayappan, and C.~Tseng.
\newblock {UTS}: An unbalanced tree search benchmark.
\newblock {\em Languages and Compilers for Parallel Computing}, pages 235--250,
  2007.

\bibitem{PapadimitriouSteiglitz82}
C.~H. Papadimitriou and K.~Steiglitz.
\newblock {\em Combinatorial Optimization: Algorithms and Complexity}.
\newblock Prentice-Hall, 1982.

\bibitem{Sanders95}
P.~Sanders.
\newblock Fast priority queues for parallel branch-and-bound.
\newblock In {\em Parallel Algorithms for Irregularly Structured Problems
  ({IRREGULAR})}, volume 980 of {\em {L}ecture {N}otes in {C}omputer
  {S}cience}, pages 379--393, 1995.

\bibitem{Song09}
F.~Song, A.~YarKhan, and J.~Dongarra.
\newblock Dynamic task scheduling for linear algebra algorithms on
  distributed-memory multicore systems.
\newblock In {\em High Performance Computing Networking, Storage and Analysis
  {(SC)}}, pages 1--11, 2009.

\bibitem{Squillante93}
M.~Squillante and E.~Lazowska.
\newblock Using processor-cache affinity information in shared-memory
  multiprocessor scheduling.
\newblock {\em {IEEE} Transactions on Parallel and Distributed Systems},
  4(2):131 --143, 1993.

\bibitem{Weissman98}
B.~Weissman.
\newblock Performance counters and state sharing annotations: a unified
  approach to thread locality.
\newblock In {\em 8th International Conference on Architectural Support for
  Programming Languages and Operating Systems {(ASPLOS)}}, pages 127--138,
  1998.

\bibitem{ppopp13}
M.~Wimmer, D.~Cederman, J.~L. Tr{\"a}ff, and P.~Tsigas.
\newblock Work-stealing with configurable scheduling strategies.
\newblock In {\em 18th ACM Symposium on Principles \& Practice of Parallel
  Programming {(PPoPP)}}, pages 315--316, 2013.

\bibitem{mtaap13}
M.~Wimmer, M.~P\"oter, and J.~L. Tr\"aff.
\newblock The {Pheet} task-scheduling framework on the {Intel}\textregistered
  {Xeon} {Phi}\texttrademark coprocessor and other multicore architectures.
\newblock In {\em Workshop on Multi-threaded Architectures and Applications
  {(MTAAP)} at the 26th International Parallel and Distributed Processing
  Symposium {(IPDPS)}}, 2013.

\end{thebibliography}

\end{document}